\documentclass[preprint,aps,amsmath,amssymb,12pt]{revtex4}

\usepackage{epsfig}
\usepackage{slashed}
\usepackage{graphicx}
\usepackage{multirow,color}
\usepackage{amsmath}
\usepackage{float}
\usepackage{diagbox}
\usepackage{epstopdf}
\usepackage{CJK}
\usepackage{color}
\usepackage{xcolor}
\usepackage{times}
\usepackage{subfigure}
\usepackage{bm}
\usepackage{braket}
\usepackage{booktabs}
\usepackage{array}
\usepackage[mathscr]{euscript}
\usepackage{caption}
\usepackage{makecell}
\usepackage{hyperref}

\makeatletter

\newcommand{\Rmnum}[1]{\expandafter\@slowromancap\romannumeral #1@}
\makeatother
\graphicspath{{fig/}}
\begin{document}
\title{ Lepton flavour violation Signals of the singly charged scalar singlet at the ILC}
\author{Chong-Xing Yue$^{1,2}$}
\thanks{cxyue@lnnu.edu.cn}
\author{Xiao-Chen Sun$^{1,2}$}
\thanks{xcsun0315@163.com}

\author{Na-Qian Zhang$^{1,2}$}
\thanks{nqzhang0610@163.com}
\author{Yang-Yang Bu$^{1,2}$}
\thanks{byy20011020@163.com}

\affiliation{
$^1$Department of Physics, Liaoning Normal University, Dalian 116029, China\\
$^2$Center for Theoretical and Experimental High Energy Physics, Liaoning Normal University, China
}

\begin{abstract}
The singly charged $SU(2)_L$ singlet scalar is one of the very interesting new particles, as it can generate neutrino masses at loop level, produce contributions to various flavour observables. We study the possibility of detecting this kind of scalar predicted by the singly-charged scalar model at ILC via the lepton flavour violation (LFV) process $e^+e^-\rightarrow S^+S^-\rightarrow \mu e  + {E\mkern-10.5 mu/}$.  Considering the constraints on the free parameters, we obtain the expected sensitivities of the ILC with the center of mass energy $\sqrt{s}=1~\mathrm{TeV}$ and the integrated luminosity $\mathcal{L}=$ $1.5~\mathrm{ab}^{-1}$ to the parameter space of the singly-charged scalar model. The prospective excluded mass range at $95\%$ C.L. is $M_S \gtrsim 470~\mathrm{GeV}$, $410~\mathrm{GeV}$ for the branching ratio $\mathcal{B}_{\mu e}$ = $100\%$ , $50\%$, respectively, while the scalar with $M_S \gtrsim 300~\mathrm{GeV}$ is excluded at $95\%$ C.L. for $\mathcal{B}_{\mu e}$ = $30\%$.
\end{abstract}

\maketitle

\section{Introduction}

The Standard Model (SM) of particle physics is the most successful theoretical description of the observed fundamental  particles up to date. Nevertheless, it is widely considered to be a low-energy effective theory  since there are still some discrepancies between experimental measurements and SM predictions. Therefore, some new physics (NP) beyond the SM are needed to explain them. The charged scalar is certainly a new physics signal, as predicted in many extensions of the SM. The double charged scalars have already, currently, been significantly discussed in many new physics models, such as the Georgi-Machacek model \cite{MAGG198061} and the Type-II seesaw model \cite{PhysRevD.22.2860,PhysRevD.22.2227,LAZARIDES1981287,PhysRevD.23.165,Kang:2014lwn}, and one of the most prominent examples is the charged Higgs in the Two Higgs Doublet Model (THDM) \cite{Branco:2011iw,GEORGI1985463}. The discussion of the triplet scalars can be found in Refs. \cite{FileviezPerez:2008jbu,Chao:2008mq,Han:2015hba,PhysRevD.8.1226}.

The singly charged $SU(2)_L$ singlet scalar is a very interesting new particle, as it can generate neutrino masses at loop level, produce contributions to various
flavour observables. Since the singly-charged scalar cannot couple to quarks, it is experimentally weakly constrained by other processes involving quarks. Furthermore, Ref. \cite{Crivellin:2020klg} has shown that the Cabibbo Angle Anomaly (CAA) and the hints of the violation of Lepton Favour Universality (LFU) in $\tau$ decays, overall called as the "flavour anomalies", can be simultaneously explained with a singly-charged scalar singlet.

The singly charged scalar singlet has been predicted in a number of models, including the well-known Zee model \cite{ZEE1980389,WOLFENSTEIN198093,Babu:2019mfe} whose phenomenology has discussed in detail in \cite{Herrero-Garcia:2017xdu}, as well as the Zee-Babu model \cite{ZEE1985141,ZEE198699,BABU1988132}, whose phenomenology has been exploited in \cite{Nebot:2007bc,Ohlsson:2009vk,Herrero-Garcia:2014hfa}.
This kind of new scalar particles can be pair-produced through the s-channel Drell-Yan process mediated by either $\gamma$ or $Z$ boson and the t-channel processes mediated by a light neutrino at $e^+e^-$ colliders, and the pair-production channel at the Large Hadron Collider (LHC) is only the s-channel Drell-Yan process. Reference \cite{Babu:2019mfe} has shown that various LEP and LHC searches can give strict constraints on the light charged scalar. However, as stated in \cite{Herrero-Garcia:2017xdu}, the mass of the new scalars should be at most a few$~\mathrm{TeV}$, meaning that charged scalars, especially those produced in pairs by the Drell-Yan process, can be detected at high energy collider experiments.  Results for new charged scalars with masses below $2~\mathrm{TeV}$ are presented in \cite{Herrero-Garcia:2014hfa}. References \cite{Nebot:2007bc,AristizabalSierra:2006gb} have shown that larger mass scales are always allowed since there is no significant deviation from the SM besides the neutrino mass.

We will focus our attention on the singly charged $SU(2)_L$ singlet scalar ($S^{\pm}$), which is the main origin of neutrino masses and interacts only with the left-handed SM lepton doublets via an antisymmetric
Yukawa coupling at tree level \cite{Felkl:2021qdn}. In this scenario simply called the singly-charged scalar model in this paper, the singly charged scalar singlet may be light, which governs the contributions to the observables, and the constraints on the Yukawa coupling $g^{ij}_S$ are given by the measured mixing and mass hierarchy of neutrinos. The singly-charged scalar model can potentially explain the flavour anomalies and generate rich phenomenology in current or future collider experiments. In this work, we will consider the feasibility of testing  the singly-charged $SU(2)_L$ singlet scalar at the ILC  with the center of mass energy $\sqrt{s}=1~\mathrm{TeV}$ and the integrated luminosity $\mathcal{L}=$ $1.5~\mathrm{ab}^{-1}$ \cite{ILC:2007bjz,Phinney:2007gp,CLICPhysicsWorkingGroup:2004qvu} and focus on its lepton flavour violation (LFV) signals induced by the process $e^+e^-\rightarrow S^+S^-\rightarrow \mu e + {E\mkern-10.5 mu/}$ with $\mu e$ being $\mu^+ e^-$ or $\mu^- e^+$ throughout the paper.

The remaining sections of the paper are arranged as follows: in section \ref{sec:Model}, we sketch singly-charged scalar model, describe the couplings and decays of the  singly charged scalar singlet and discuss the bounds on the relevant free parameters. In section \ref{sec:Signature}, we present the simulation results for the LFV signal process $e^+e^-\rightarrow S^+S^-\rightarrow \mu e + {E\mkern-10.5 mu/}$. Section \ref{sec:Conclusions and discussions} contains our conclusions and simple discussions.
\section{Model Description \label{sec:Model}}
An extension of the SM by charged scalar particle being singlet under the $SU(2)_{L}$ gauge group is called as the singly-charged scalar model \cite{Felkl:2021qdn}, in which the  singly charged scalar singlet $S$ is regarded as the main origin of the neutrino masses and makes at least one of the external neutrinos coupled to two SM left-handed lepton doublets via the antisymmetric Yukawa coupling. The Lagrangian pertaining to the singly charged scalar singlet $S$ is given by \cite{Felkl:2021qdn}

\begin{eqnarray}\label{kinetic_Lagrangian}
 \mathcal{L}_S = -S^*(D^\mu D_\mu + M^2_S)S + (g^{ij}_S\bar{L}^C_{i}\,L_jS + h.c.).
 \end{eqnarray}

The first sector is the kinetic term,  after breaking electroweak symmetry, which generates couplings to the photon and the $Z$ boson at the tree level, but not to the $W^\pm$ bosons.
The second part describes Yukawa-type interactions with SM leptons that entail significant flavour violations for charged leptons.  $M_S$ is the mass of the singly charged scalar singlet $S$, $L_i \equiv (\nu_i,\ell_i)^T$ denotes the SM left-handed lepton doublet, $C$ represents charge conjugation operation, $i$ and $j$ are flavour indices which are desired to be summed up when repeated. The Yukawa coupling $g^{ij}_S$ is antisymmetric, $g^{ij}_S=-g^{ji}_S$, and the $3\times3$ Yukawa coupling matrix $g_S$ can be written as

\begin{eqnarray}
    g_S =
    \left(
    \begin{array}{ccc}
        0 & g^{e\mu}_S & g^{e\tau}_S \\
        -g^{e\mu}_S & 0 & g^{\mu\tau}_S \\
        -g^{e\tau}_S & -g^{\mu\tau}_S & 0 \\
    \end{array}
    \right).
\end{eqnarray}

After electroweak symmetry breaking, the second part of the Lagrangian, Eq.(1), for the Yukawa interactions with leptons is stated as

\begin{eqnarray}
\mathcal{L}_S^{lept} = 2 [g^{e\mu}_S(\bar{\nu}^c_{e}\mu_L-\bar{\nu}^c_{\mu}e_L) + g^{e\tau}_S(\bar{\nu}^c_{e}\tau_L
-\bar{\nu}^c_{\tau}e_L) \\ ~~~~~~~~~+ g^{\mu\tau}_S( \bar{\nu}^c_{\mu}\tau_L - \bar{\nu}^c_{\tau}\mu_L)] S + h.c.\nonumber
\end{eqnarray}
so there are three free parameters in the singly-charged scalar model which are $g^{e\mu}_S$, $g^{e\tau}_S$, and $g^{\mu\tau}_S$.
However, they can be linked through the measured mixing and mass hierarchy of neutrinos and can be turned to one free parameter via the following relations\cite{Felkl:2021qdn}:
\begin{eqnarray}
    \frac{g^{e\tau}_S}{g^{\mu\tau}_S} & = -\frac{\sin(\theta_{23})}{\tan(\theta_{13})}e^{i\delta}, \\
    \frac{g^{e\mu}_S}{g^{\mu\tau}_S} & = \frac{\cos(\theta_{23})}{\tan(\theta_{13})}e^{i\delta},
\end{eqnarray}
for the case of Inverted Ordering (IO) $m_3 \leq m_1 < m_2$ of neutrino masses and
\begin{eqnarray}
    \frac{g^{e\tau}_S}{g^{\mu\tau}_S} & = \tan(\theta_{12})\frac{\cos(\theta_{23})}{\cos(\theta_{13})} + \tan(\theta_{13})\sin(\theta_{23})e^{i\delta}, \\
    \frac{g^{e\mu}_S}{g^{\mu\tau}_S} & = \tan(\theta_{12})\frac{\sin(\theta_{23})}{\cos(\theta_{13})} - \tan(\theta_{13})\cos(\theta_{23})e^{i\delta},
\end{eqnarray}
for Normal Ordering (NO) $m_1 < m_2 \leq m_3$ . Using the data in TABLE~\ref{Tab:NuFIT} we can obtain: $g^{e\mu}_S \sim g^{e\tau}_S \sim 4 g^{\mu\tau}_S$ for IO , and $g^{\mu\tau}_S \sim 1.5 g^{e\mu}_S \sim 3 g^{e\tau}_S$ for NO.

\begin{table}[htb]
	\centering{
\caption{\label{Tab:NuFIT} The experimental values for leptonic mixing parameters taken from NuFIT 5.0~\cite{Esteban:2020cvm}. }

		\begin{tabular}{c|c c c c}
			\hline
                & $\delta$\;[rad] & $\sin^2(\theta_{12})$ & $\sin^2(\theta_{13})$ & $\sin^2(\theta_{23})$ \\
                \hline
                IO & $4.92\pm0.45$ & $0.304\pm0.012$ & $0.02238\pm0.00062$ & $0.575\pm0.016$ \\
                NO & $3.44\pm0.42$ & $0.304\pm0.012$ & $0.02219\pm0.00062$ & $0.573\pm0.016$ \\
              \hline
	\end{tabular}}	
\end{table}

From Eq.(3) we can see that the singly charged scalar singlet $S$ can decay into a neutrino $\nu$ and a charged lepton $\ell$, in the case of ignoring the charged lepton mass, the partial width is given by

\begin{eqnarray}
    \Gamma(S\to \ell_i\nu_j) = \Gamma(S\to \ell_j\nu_i) = \frac{|g^{ij}_S|^2}{4\pi}M_S\,.
\end{eqnarray}

If the neutrino flavour is not identified, the branching ratio can be written as

\begin{eqnarray}
    \mathcal{B}(S\to \ell_i\nu_j) = \frac{\sum_{j\neq i} |g^{ij}_S|^2}{2(|g_S^{e\mu}|^2+|g_S^{e\tau}|^2+|g_S^{\mu\tau}|^2)}\,.
\end{eqnarray}
Thus, for the IO and NO cases, the branching ratio $\mathcal{B}_{\mu e}$ of the $\mu e$ channel is approximately $50\%$ and $30\%$, respectively. We can also get the branching ratios of the $e\tau$ and $\mu\tau$ channels.
Although the singly charged scalar singlet $S$ has six decay channels, there are only three different signals that can be observed in the case that neutrinos are indistinguishable in experiments.

The contributions of the singly-charged scalar model to the low-energy observables  are  dominated by the singly charged scalar singlet $S$. The relevant experimental data can generate constraints on the free parameters  $M_S$ and $g^{ij}_S$. For example, Singly-charged scalar singlet can induce leptonic non-standard neutrino interactions (NSIs) at the tree level \cite{Ohlsson:2009vk}. After integrating out the singly-charged scalar singlet, the low-energy effective four-fermion interactions can be written as
\begin{eqnarray}\label{eqn:ncl}
    \mathcal{L}^{NSI}_{d=6} = -2\sqrt{2}G_F\varepsilon^{kl}_{ij}\left(\nu^\dagger_i\gamma^\mu P_L \nu_j\right)\left(\ell^\dagger_k\gamma_{\mu} P_L \ell_l\right).\
\end{eqnarray}
 The effective NSI parameters can be expressed as
\begin{eqnarray}\label{eq:NSI}
    \varepsilon_{ij}^{kl} = -\frac{1}{\sqrt{2}G_F}\frac{(g_S^{ik})^*g_S^{jl}}{M_S^2}.
\end{eqnarray}
Although NSI has not been confirmed experimentally, its effects have been extensively studied in various new physics scenarios and the stringent bounds from various
experiments are obtained, for example see Ref. \cite{Proceedings:2019qno} for review. If one assumes that the upper limits on  the  leptonic NSI parameters are  $|\varepsilon_{ij}^{kl}|\sim\ 10^{-2}$, then a constraint $|g^{e\mu}_S|/ M_S\sim1.28\times10^{-1}/\mathrm{TeV}$ can be  obtained from above equation.

Singly-charged scalar singlet may affect the partial decay width $\Gamma_{i\to j}$ associated to the decay process $\ell_i\to \ell_j\nu\nu$
\begin{eqnarray} \label{CAAvarepsilon}
    \delta(\ell_i\to \ell_j\nu\nu) \equiv \frac{1}{\sqrt{2}G_F}\frac{|g^{ij}_S|^2}{M^2_S}.
\end{eqnarray}
For $\delta(\mu\to e\nu\nu)= 0.00065(15)$~\cite{Crivellin:2020klg}, we can obtain $|g^{e\mu}_S| /M _S \approx 1.03\times 10^{-1}/~\mathrm{TeV}$. Singly-charged scalar singlet can also lead to negative correction to the $W$ boson mass~\cite{Crivellin:2020klg},
\begin{eqnarray}
    \delta M^2_W = -\frac{M^2_W}{\sqrt{2}G_F}\left|1 - \frac{M_WM_Z}{2M^2_W - M^2_Z}\right|\frac{|g^{e\mu}_S|^2}{M^2_S}.
\end{eqnarray}
If we allow for a $3\sigma$ difference among the SM prediction and the world average of the measurements~\cite{Erler:2019ddu, Freitas:2020kcn, ParticleDataGroup:2020ssz}, then there is the constraint $|g^{e\mu}_S|/M_S \approx 1.12\times 10^{-1}/~\mathrm{TeV}$.

The mass $M_S$ is regarded as another free parameter.  Ref. \cite{Crivellin:2020klg} has given a model-independent lower bound $M_S \geq 200~\mathrm{GeV}$. Using the LEP and LHC data, Ref. \cite{Cao:2017ffm} has shown that singly-charged scalar singlet with mass above $65~\mathrm{GeV}$ is still allowed. So, in this paper, we take $M_S \geq 300~\mathrm{GeV}$ which could be studied safely beyond any mass region that might be excluded.

\section{Signatures of the Singly-Charged Scalar Singlet at the ILC \label{sec:Signature}}

\begin{figure}[htb]
\begin{center}
\centering\includegraphics [scale=0.5] {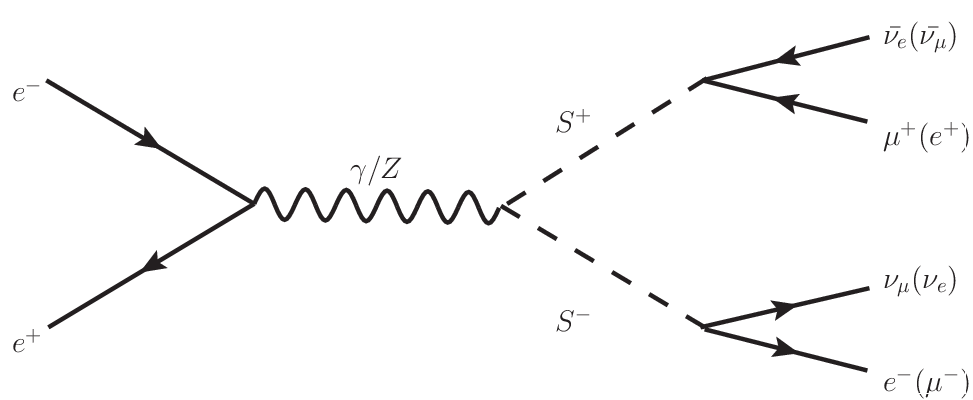}
\caption{The Feynman diagram for the process of $e^+e^-\rightarrow S^+S^-\rightarrow \mu^+e^- + {E\mkern-10.5 mu/}$.}
\label{HSaZ}
\end{center}
\end{figure}

In this subsection we concern the potential on searching for the LFV signals of the  singly charged $SU(2)_L$ singlet scalar $S$ via its  pair production at the $1~\mathrm{TeV}$ ILC with integrated luminosity of $\mathcal{L}=$ $1.5~\mathrm{ab}^{-1}$. The scalar $S$ can be generated in pairs via s-channel Drell-Yan processes which are mediated by either $\gamma$ or $Z$ boson and can also be produced in pairs via t-channel processes
which are mediated by light neutrinos. Since the contributions of the t-channel processes are much smaller than those of the s-channel Drell-Yan processes, we  only consider the s-channel Drell-Yan
processes in this paper.
It is well known that the lepton $\tau$ in final state decays quickly and complex, making the signal difficult to identify. So, for simplicity of analysis, the signals involving lepton $\tau$ are neglected, and only the $\mu e$ signals are considered. We will discuss the LFV signals of the  singly charged $SU(2)_L$ singlet scalar $S$ in three scenarios. The first scenario is assumed $g_S^{e\tau}=g_S^{\mu\tau}=0$, i.e., the branching ratio $\mathcal{B}_{\mu e}$ of the $\mu e$ decay channel is 100\%.
The second and third cases we considered corresponding to $\mathcal{B}_{\mu e} = 50\%$ (IO) and $\mathcal{B}_{\mu e}$ = $30\%$ (NO) , respectively.
The Feynman diagrams for the $s$-channel Drell-Yan process $e^+e^-\rightarrow S^+S^-\rightarrow \mu e + {E\mkern-10.5 mu/}$ with $\mu e$ being $\mu^+ e^-$ or $\mu^- e^+$ are shown in Fig.~\ref{HSaZ}. The signal of $S^\pm$ will be identified through the detection of $\mu e$ plus missing energy corresponding to the SM neutrinos $\nu_e$ and $\nu_\mu$. The main SM backgrounds considered in our numerical analysis are

\begin{itemize}
\item {$e^+e^-\rightarrow \mu e + {E\mkern-10.5 mu/}$},
\item {$e^+e^-\rightarrow l l l l$ ($l=e, \mu, \tau$) },
\item {$e^+e^-\rightarrow H  \nu_e  \bar{\nu_e}(H\rightarrow W^+W^- \rightarrow \mu e + {E\mkern-10.5 mu/}$)},
\item {$e^+e^-\rightarrow H  Z(H\rightarrow W^+W^- \rightarrow \mu e + {E\mkern-10.5 mu/}, Z\rightarrow \nu_l  \bar{\nu_l})$},
\item {$e^+e^- \rightarrow \tau^+\tau^- \rightarrow \mu e + {E\mkern-10.5 mu/}$ },
\item {$e^+e^-\rightarrow W^+W^-\rightarrow \tau^+\tau^-   \nu_\tau  \bar{\nu_\tau}$ ($\tau^+$ and $\tau^-$ decay leptonically)}.
\end{itemize}
The most dominant background comes from the process $e^+e^-\rightarrow \mu e + {E\mkern-10.5 mu/}$. For the process $e^+e^-\rightarrow l l l l$, if two of leptons in the final state escape detection being regarded as ${E\mkern-10.5 mu/}$, then it can also contribute to the SM background.
The missing energy ${E\mkern-10.5 mu/}$ contributed by any of the neutrino final states which also might include the background events produced from the leptonic decay of $W^{\pm}$ gauge bosons.

We use the package \texttt{FeynRules}~\cite{Alloul:2013bka} to generate the UFO model file~\cite{Degrande:2011ua} for the singly charged scalar singlet $S$. For the numerical results and the event generation, we choose the Monte-Carlo (MC) simulation with the \texttt{MadGraph5\_aMC@NLO}~\cite{Alwall:2014hca} to generate the signal and background events at parton level with basic cuts: $p_T^{\ell} > 10~\mathrm{GeV}$, $|\eta^\ell | < 2.5$, which refer to the transverse momentum and the pseudorapidity of the electron and muon, respectively. Furthermore, we employ separately the packages \texttt{Pythia8}~\cite{Sjostrand:2014zea} and \texttt{Delphes-3.5.1}~\cite{deFavereau:2013fsa} for parton shower and the delphes-card-ILD.tcl~\cite{Behnke:2013lya} for detector simulation.

\begin{figure}[htb]
\begin{center}
\centering\includegraphics [scale=0.4] {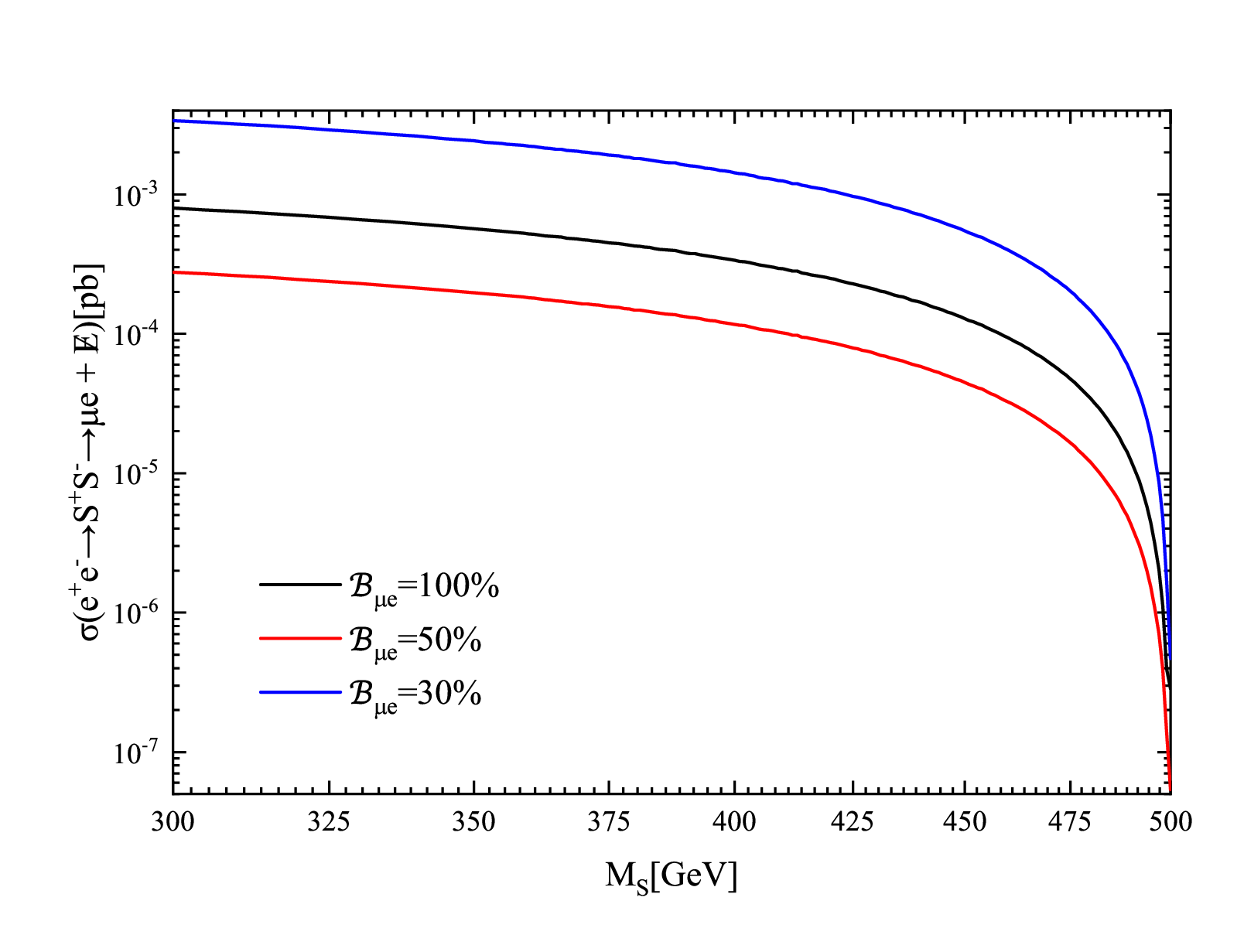}
\caption{The production cross section of the LFV signal process $e^+e^-\rightarrow S^+S^-\rightarrow \mu e + {E\mkern-10.5 mu/}$ as a function of $M_S$ at the $1~\mathrm{TeV}$ ILC.}
\label{cross}
\end{center}
\end{figure}

The production cross section of the LFV signal process $e^+e^-\rightarrow S^+S^-\rightarrow \mu e + {E\mkern-10.5 mu/}$ at the $1~\mathrm{TeV}$ ILC is plotted in Fig.~\ref{cross} as a function of the mass parameter $M_S$, where the different curves correspond the cross sections for the values of the branching ratio $\mathcal{B}_{\mu e}$ as $100\%$, $50\%$, $30\%$. We can see that the production cross section can  respectively reach $3.43\times10^{-3}$ pb, $7.98\times10^{-4}$ pb and $2.77\times10^{-4}$ pb for the values of $\mathcal{B}_{\mu e}$ as $100\%$, $50\%$ and $30\%$. The signal cross section is smaller than the corresponding SM background cross section ($0.09818$ pb).

\begin{figure}[htb]
\begin{center}
\subfigure[]{\includegraphics [scale=0.3] {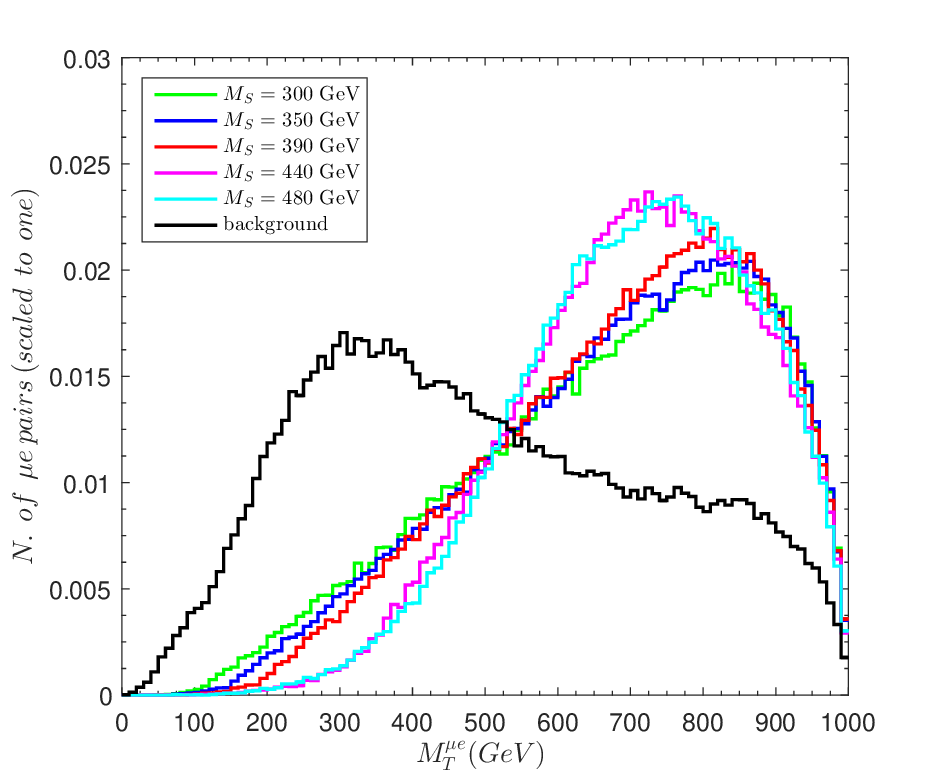}}
\subfigure[]{\includegraphics [scale=0.3] {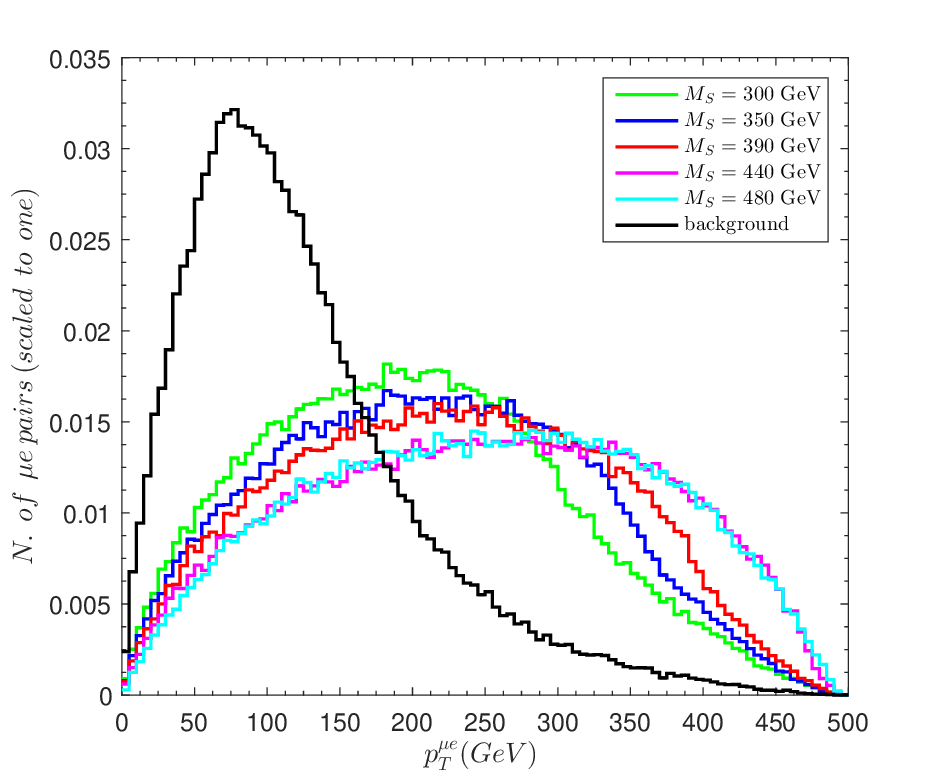}}
\subfigure[]{\includegraphics [scale=0.3] {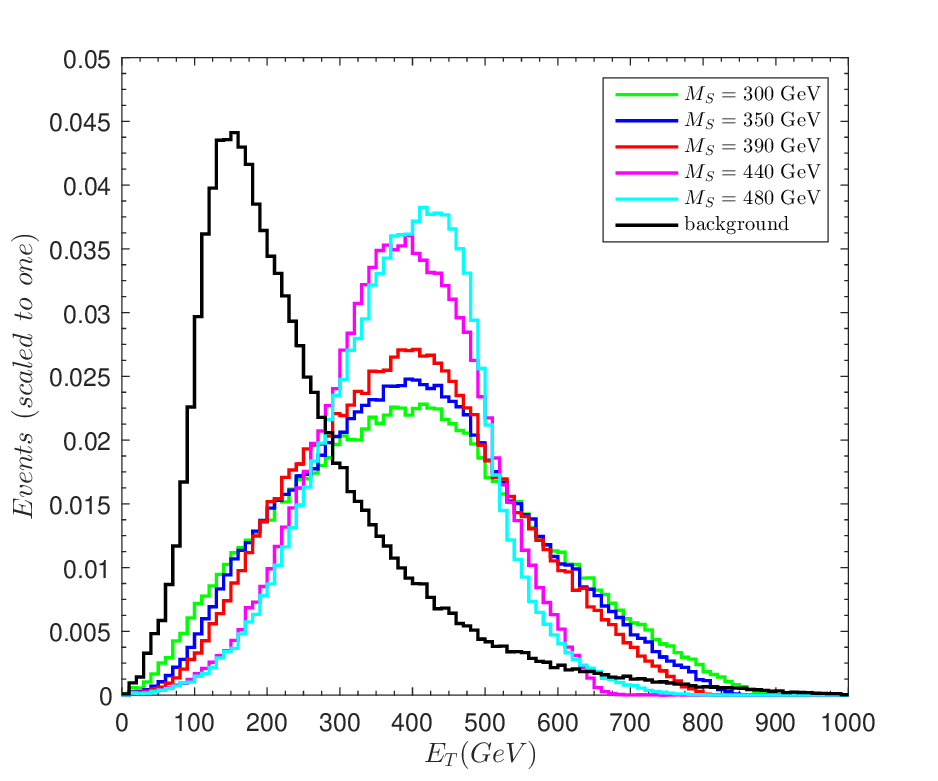}}
\caption{Normalized distributions of $M^{\mu e}_T$ (a), $p^{\mu e}_T$ (b), $E_T$ (c) from the signal and background events for different $M_S$ benchmark points at the ILC with $\sqrt{s}=1~\mathrm{TeV}$ and $\mathcal{L}=$ $1.5~\mathrm{ab}^{-1}$.}
\label{Normalized distributions}
\end{center}
\end{figure}

The criteria used to reject the background is based on the full kinematic study of the events. For that, we have examined the kinematic distributions of the signal and background events to define a convenient set of cuts that can provide a good discrimination against the background. We perform the analysis of the resulting output by \texttt{MadAnalysis5}~\cite{Conte:2012fm,Conte:2014zja,Conte:2018vmg}. In Fig.~\ref{Normalized distributions}, we show the normalized distributions of the transverse mass of the system comprised of $\mu e$ and the missing momentum: $M^{\mu e}_T$, the transverse momentum: $p^{\mu e}_T$, and the total transverse energy: $E_T$, for the background and signal events at typical mass points of $M_S= 300~\mathrm{GeV}$, $350~\mathrm{GeV}$, $390~\mathrm{GeV}$,  $440~\mathrm{GeV}$, $480~\mathrm{GeV}$ at the $1~\mathrm{TeV}$ ILC with $\mathcal{L}=$ $1.5~\mathrm{ab}^{-1}$.

From Fig.3, we can see that the distributions of signal in the mass range considered in this paper have good distinguish from the background. According to the information of these kinematic distributions, different optimized cuts as presented in TABLE~\ref{cut} can apply to suppress background and increase statistically significance.
Particularly, the transverse momentum $p^{\mu e}_T$ of the background is distributed around the $W$ boson mass as shown in Fig.~\ref{Normalized distributions} (b). This is because the main contribution to the transverse momentum of the background event originates from the leptonic decay of the $W^{\pm}$ gauge bosons. So we take the cut $p^{\mu e}_T > 175~\mathrm{GeV}$ to reduce the background.
As shown in FIG.~\ref{Normalized distributions} (a) and (c), after applying the selection cuts, the transverse mass $M^{\mu e}_T$ and total transverse energy $E_T$ in the background event are softer than those in the signal event.
In principle, the other variables are available to us to discriminate the signal from the background. Nevertheless, such variables are quite similar and their effects would not be significantly better than the kinematic variables described above. So, we do not consider them in this paper. The cross sections after applying the above selection cuts for the signal and background is summarized in TABLE~\ref{Tab:signal1}.

\begin{table}[htb]
\centering
\caption{Selected cuts on signal and background for $300~\mathrm{GeV}\leq M_S\leq500~\mathrm{GeV}$ .}
\label{cut}
\begin{tabular}{c | c}
\hline\hline
\multirow{1}{*}{Cuts}  & \multicolumn{1}{c}{$300~\mathrm{GeV}\leq M_S\leq500~\mathrm{GeV}$}\\
\hline
		 \makecell{Cut1:the transverse mass of the system comprised\\ of the objective and the missing momentum} & $M^{\mu e}_T  > 525~\mathrm{GeV}$    \\
\hline
		Cut2: the transverse momentum & $p^{\mu e}_T > 175~\mathrm{GeV}$ \\
\hline
		Cut3: the total transverse energy & $E_T > 300~\mathrm{GeV}$   \\
\hline\hline
\end{tabular}
\end{table}

\begin{table}[htb]\tiny
	\centering{
\caption{\label{Tab:signal1} After different cuts applied, the cross sections for the signal and SM background with benchmark points for $\mathcal{B}_{\mu e}$ = $100\%$($30\%$) at the $\sqrt{s} = 1~\mathrm{TeV}$ ILC with $\mathcal{L}=$ $1.5~\mathrm{ab}^{-1}$.}
		\begin{tabular}{m{1.3cm}<{\centering}|m{2.0cm}<{\centering} m{2.0cm}<{\centering} m{2.0cm}<{\centering} m{2.0cm}<{\centering} m{2.0cm}<{\centering} |m{2.0cm}<{\centering}}
			\hline\hline
     \multirow{2}{*}{Cuts} & \multicolumn{5}{c|}{signal [pb] $\mathcal{B}_{\mu e}$ = $100\%$($30\%$)} & \multirow{2}{*}{background [pb]} \\
     \cline{2-6}
     & $M_S=300~\mathrm{GeV}$
     & $M_S=350~\mathrm{GeV}$
     & $M_S=390~\mathrm{GeV}$
     & $M_S=440~\mathrm{GeV}$
     & $M_S=480~\mathrm{GeV}$  & \\ \hline
     Basic Cuts
                 & \makecell{$3.43\times10^{-3}$\\$(2.77\times10^{-4} )$}
                 & \makecell{$2.42\times10^{-3}$\\$(1.97\times10^{-4} )$}
                 & \makecell{$1.62\times10^{-3}$\\$(1.32\times10^{-4} )$}
                 & \makecell{$7.14\times10^{-4}$\\$(5.83\times10^{-5} )$}
                 & \makecell{$1.45\times10^{-4}$\\$(1.19\times10^{-5} )$}
                 & $9.82\times10^{-2}$
        \\
     Cut 1
                 & \makecell{$1.49\times10^{-3}$\\$(1.22\times10^{-4} )$}
                 & \makecell{$1.17\times10^{-3}$\\$(9.45\times10^{-5} )$}
                 & \makecell{$7.86\times10^{-4}$\\$(6.46\times10^{-5} )$}
                 & \makecell{$3.74\times10^{-4}$\\$(3.07\times10^{-5} )$}
                 & \makecell{$7.83\times10^{-5}$\\$(6.67\times10^{-6} )$}
                 & $1.50\times10^{-2}$
       \\
     Cut 2
                 & \makecell{$1.34\times10^{-3}$\\$(1.10\times10^{-4} )$}
                 & \makecell{$1.07\times10^{-3}$\\$(8.59\times10^{-5} )$}
                 & \makecell{$7.27\times10^{-4}$\\$(5.95\times10^{-5} )$}
                 & \makecell{$3.55\times10^{-4}$\\$(2.92\times10^{-5} )$}
                 & \makecell{$7.55\times10^{-5}$\\$(6.45\times10^{-6} )$}
                 & $1.01\times10^{-2}$
       \\
     Cut 3
                 & \makecell{$1.23\times10^{-3}$\\$(1.00\times10^{-4} )$}
                 & \makecell{$9.73\times10^{-4}$\\$(7.77\times10^{-5} )$}
                 & \makecell{$6.55\times10^{-4}$\\$(5.38\times10^{-5} )$}
                 & \makecell{$3.21\times10^{-4}$\\$(2.64\times10^{-5} )$}
                 & \makecell{$6.95\times10^{-5}$\\$(5.89\times10^{-6} )$}
                 & $6.51\times10^{-3}$
     \\ \hline
     $SS$     & $17.34$ $(1.53 )$ & $14.11$ $(1.24 )$  &  $9.60$ $(0.86 )$  &  $4.18$ $(0.36 )$  &  $0.71$ $(0.05 )$ \\ \hline\hline
	\end{tabular}}	
\end{table}

It can be seen that the background is effectively suppressed. The values of the statistical significance $SS=S/\sqrt{S + B}$ obtained with the selection strategy are listed in the last row, where $S$ and $B$ are the number of events for the signal and background. The values of the statistical significance $SS$ can approximately reach $4.18$ and $0.71$ at $M_S=440~\mathrm{GeV}$ and $480~\mathrm{GeV}$ for $\mathcal{B}_{\mu e}$ = $100\%$.

\begin{figure}[htb]
\begin{center}
\centering\includegraphics [scale=0.41] {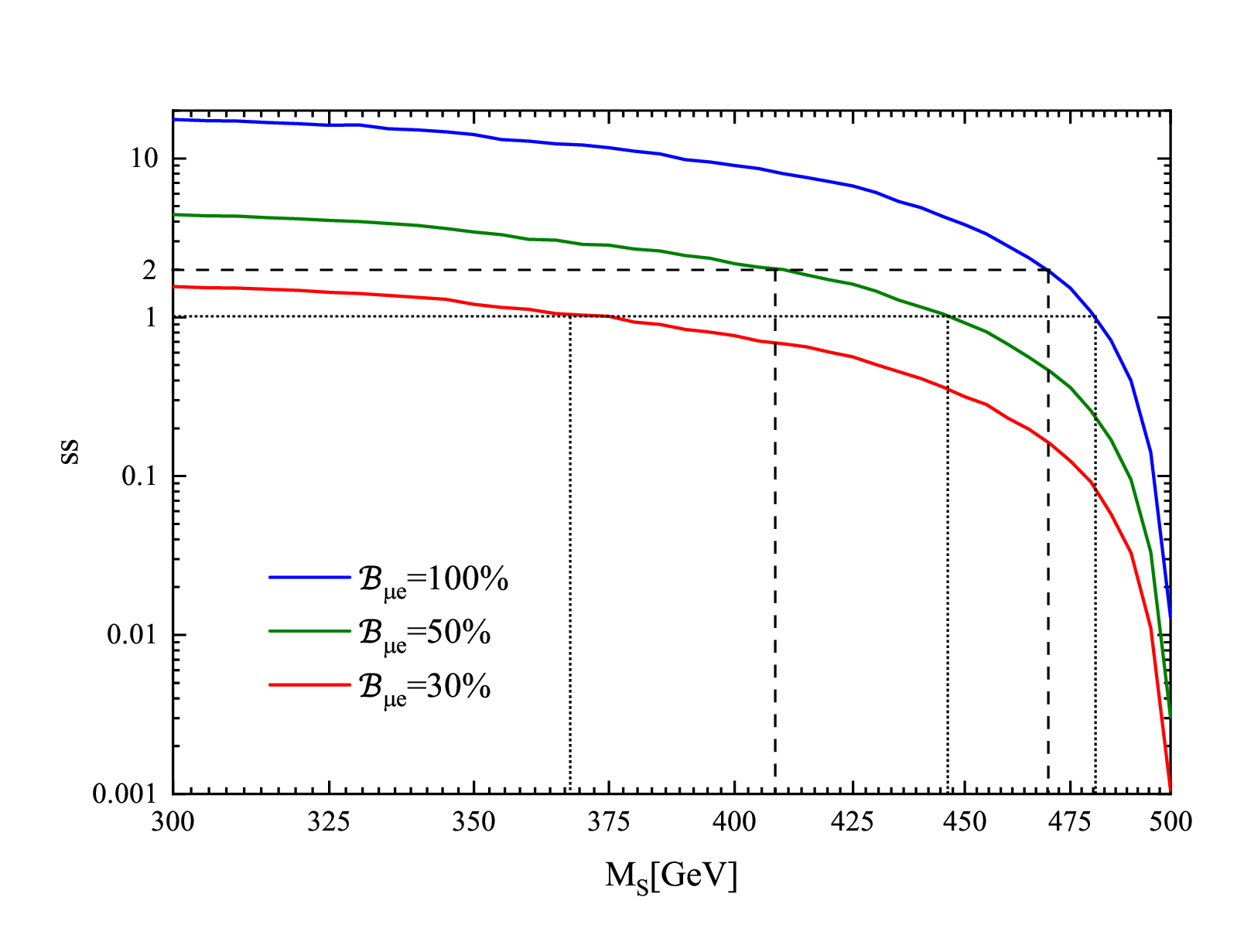}
\caption{The SS curves for the LFV process $e^+e^-\rightarrow S^+S^-\rightarrow \mu e + {E\mkern-10.5 mu/}$ at the ILC with $\sqrt{s}=1~\mathrm{TeV}$ and $\mathcal{L}=$ $1.5~\mathrm{ab}^{-1}$.}
\label{235sigma}
\end{center}
\end{figure}

Based on the above benchmark points, the expected exclusion bounds of the free parameters $M_S$ at the ILC with $\sqrt{s}=1~\mathrm{TeV}$ and $\mathcal{L}=$ $1.5~\mathrm{ab}^{-1}$ are shown in Fig.~\ref{235sigma}, where $100\%$, $50\%$ and $30\%$ are shown in blue, green and red lines, respectively.
From Fig.~\ref{235sigma} we can see the values of the statistical significance $SS$ are about $[17.34,0.01]$, $[4.31,0.003]$ and $[1.53,0.001]$ for $M_S\in[300,500]~\mathrm{GeV}$, $\mathcal{B}_{\mu e}$ = $100\%$, $50\%$ and $30\%$, respectively. As long as the values of the mass parameter $M_S$ are larger than  respectively $470~\mathrm{GeV}$ and $410~\mathrm{GeV}$  for $\mathcal{B}_{\mu e}$ = $100\%$ and $50\%$, the LFV signals of the  singly charged $SU(2)_L$ singlet scalar $S$ might be excluded at $95\%$ C.L. by the $1~\mathrm{TeV}$ ILC with $\mathcal{L}=$ $1.5~\mathrm{ab}^{-1}$, while the scalar with $M_S \gtrsim 300~\mathrm{GeV}$ is excluded at $95\%$ C.L. for the case $\mathcal{B}_{\mu e}$ = $30\%$.
Therefore, the signal of singly-charged scalar particle $S$ might be investigated through the LFV process $e^+e^-\rightarrow S^+S^-\rightarrow \mu e + {E\mkern-10.5 mu/}$ at the ILC in near future.

\section{Conclusions and discussions \label{sec:Conclusions and discussions}}

Charged scalars exist in many new physics scenario, which can produce rich phenomenology in current or future high energy collider experiments. We study the possibility of detecting the singly charged scalar singlet $S$ predicted by the singly-charged scalar model at ILC via its pair production.  Considering the constraints on the parameter space, we focus our attention on the LFV signal of the singly charged scalar singlet $S$ generated by the process $e^+e^-\rightarrow S^+S^-\rightarrow \mu e + {E\mkern-10.5 mu/}$. The expected sensitivities of the ILC with $\sqrt{s}=1~\mathrm{TeV}$ and $\mathcal{L}=$ $1.5~\mathrm{ab}^{-1}$ to the parameter space of the singly-charged scalar model are derived.

In order to obtain the effective expected range of the free parameters of the singly charged scalar singlet $S$, we have adopted a appropriate statistical treatment to the signal and
SM background. We apply the inclusive cuts to the  variables $M^{\mu e}_T$, $p^{\mu e}_T$, $E_T$, which can  effectively suppress the SM background. The numerical results we obtained show that the expected sensitivities of the ILC with $\sqrt{s}=1~\mathrm{TeV}$ and $\mathcal{L}=$ $1.5~\mathrm{ab}^{-1}$ to the parameter space of the singly charged scalar singlet $S$ are not experimentally excluded.
The prospective excluded mass range at $95\%$ C.L. is $M_S \gtrsim 470~\mathrm{GeV}$, $410~\mathrm{GeV}$ for $\mathcal{B}_{\mu e}$ = $100\%$ , $50\%$, respectively, while the scalar with $M_S \gtrsim 300~\mathrm{GeV}$ is excluded at $95\%$ C.L. for the case $\mathcal{B}_{\mu e}$ = $30\%$. Thus, the ILC which is designed for operation at several collision energies, will has an  opportunity to search for this kind of scalar particles via the LFV process $e^+e^-\rightarrow S^+S^-\rightarrow \mu e + {E\mkern-10.5 mu/}$.

The radiative return production process at the $e^+e^-$ colliders may affect the production cross sections of some processes. The scattering of the initial radiation photons from positrons (electrons) only affects the single production of charged scalar. For the pair production of charged scalar studied in this paper, the radiative return production process can only be achieved through the initial radiation photons decay. Within the mass range of charged scalars discussed in this paper, the energy threshold for pair production process should reach more than $600~\mathrm{GeV}$, which exceeds the energy of the incoming beam. Therefore we ignore the influence of radiative return production process in our analysis.

\section*{ACKNOWLEDGMENT}

This work was partially supported by the National Natural Science Foundation of China under Grant No. 11875157 and No. 12147214.


\bibliography{work1}

\end{document}